\documentclass[]{wlscirep}
\usepackage[utf8]{inputenc}
\usepackage[T1]{fontenc}
\usepackage{bm}
\usepackage{soul}
\usepackage{multirow}

\title{The contribution of winds of star clusters to the Galactic cosmic-ray population}

\author[1,2,*]{Giada Peron}
\author[3]{Sabrina Casanova}
\author[1]{Stefano Gabici}
\author[4]{Vardan Baghmanyan}
\author[2,5,6]{Felix Aharonian}
\affil[1]{Université de Paris, CNRS, Astroparticule et Cosmologie, F-75013 Paris, France}
\affil[2]{Max Planck Institute f{\"u}r Kernphysik, Heidelberg, Germany}
\affil[3]{Institute of Nuclear Physics, Krakow, Poland}
\affil[4]{Lehrstuhl f\"{u}r Astronomie, Universit\"{a}t W\"{u}rzburg, W\"{u}rzburg, Germany}
\affil[5]{Dublin Institute for Advanced Studies, Dublin, Ireland}
\affil[6]{Yerevan State University,  1 Alek Manukyan St, Yerevan 0025, Armenia}
\affil[*]{e-mail: giada.peron@apc.in2p3.fr}

\begin{document}
\maketitle

\textbf{Cosmic rays are energetic nuclei that permeate the entire Galactic disk. Their existence requires the presence of  powerful particle accelerators. While Galactic supernova explosions may supply the required energy, there is growing evidence that they cannot explain all of the observed properties of cosmic rays, such as their maximum particle energy and isotopic composition. 
Among Galactic objects, winds from stellar clusters meet the energetic requirement and provide a suitable environment for particle acceleration. The recent detection of some of these objects in $\gamma$ rays confirms that they indeed harbor high-energy particles.
However, as most supernovae explode inside stellar clusters, it is difficult to distinguish the contribution of winds to particle acceleration. Here we report the detection of young star clusters in the nearby Vela molecular ridge star forming region. The young age of the systems guarantees an unbiased estimate of the stellar CR luminosity free from any supernova or pulsar contamination and allows us to draw conclusions on the acceleration efficiency and the total power supplied by these objects. 
We demonstrate that much more than 1\% of the wind mechanical power is converted into CRs and  {consequently} conclude that a  small but non-negligible fraction $\sim 1-10\%$ of the CR population is contributed by stellar clusters.}
\vspace{10 pt}


Clusters of massive stars are thought to be powerful particle accelerators. Their energy supply resides in the winds blown by member stars. For our Galaxy the total mechanical power injected by winds of stellar clusters (SCs) has been estimated to be  {approximately} $ 10^{41}$ erg s$^{-1}$, which is just a factor of a few smaller than the power supplied by supernova (SN) explosions\cite{Seo2018TheProduction}. 
Such a large amount of energy may support a sizeable fraction of the $\sim 7 \times 10^{40}$ erg s$^{-1}$ luminosity of Galactic cosmic rays (GCRs) \cite{Strong2010GlobalWay,Cesarsky1983a}.  {Moreover, allowing particle acceleration in stellar environment could explain the anomalous excess in the isotopic ratio $X_{CR} = ^{22}$Ne/$^{20}$Ne~$\sim 0.317$ observed in GCRs (the cosmic value being  $X_{S} \sim 0.0735$)\cite{Casse1982OnRays,Binns2005Cosmic-RayRays,Boschini2020InferenceFramework}.} 

$\gamma$-ray observations of several SCs  {from GeV to PeV energies}\cite{Aharonian2019,Abeysekara2021,Cao2021} confirmed that these objects are indeed effective particle accelerators\cite{Higdon2005OBRays,Bykov2015NonthermalSuperbubbles,Morlino2021ParticleClusters,Vieu2022CosmicSuperbubbles}.
However, the fraction of wind mechanical energy that is converted into GCRs is still unconstrained. 
This is because the $\gamma$-ray emitting clusters detected so far reside in complex regions, where several potential particle accelerators overlap: either because some SN already exploded in the cluster (this is the case for Westerlund 1\cite{Aharonian2022A1} and Cygnus OB2\cite{Wright2015TheOB2}) or because of the superposition with other potential CR sources along the line of sight (as in the case of Westerlund 2 \cite{Mestre2021Probing2}). 
This prevents an observational estimate of the contribution of SC winds to the population of GCRs. 

This problem can be solved by targeting very young SCs, i.e., younger than the typical age when SN explosions begin to occur\cite{Maurin2016EmbeddedConstraints}. Massive stars end their life with a SN explosion, and their lifetime decreases with the stellar mass ranging from $\sim$~3 Myr for the most massive stars, $\sim 120$ M$_{\odot}$, to 10 Myr for the least-massive SN progenitors, $\sim 10$ M$_{\odot}$ \cite{Ekstrom2012Grids0.014}.
 {Earlier searches in this direction resulted in upper limits on the gamma-ray flux for a handful of young star clusters \cite{Maurin2016EmbeddedConstraints}.}

Very young SCs are expected to be still embedded in the gas cocoon out of which they   {formed and where they spend 10\%-20\% of their main-sequence lifetime \cite{Wood1989MASSIVEGALAXY}}. Even though the cocoon is optically thick to the light of the stars, the presence of embedded young SCs is revealed by the near-infrared emission that dust re-emits after being heated by the strong UV radiation of massive stars. 
 Far infrared  emission (from $\sim$ 10 $\mu$m to $\sim$ 100 $\mu$m  ) acts as a  tracer of early type (O/B) stars\cite{Wood1989MASSIVEGALAXY}.  It has been demonstrated that using  FIR luminosity is equivalent as using radio to identify the regions of ionized gas (H\textsc{ii} region) around massive stars \cite{Mascoop2021TheWavelengths}. That allowed the identification of  {more than 8000 H\textsc{ii} regions in the Galaxy using their 22-$\mu$m emission \cite{Anderson2014TheRegions}.}
 {Among these,} several  {embedded} young massive SCs  {are found} in the near ($\sim$~1-2~kpc) Vela molecular cloud ridge (VMR) from which, making use of Fermi-LAT observations, we extracted the $\gamma$-ray spectrum and constrained the CR luminosity (see Table~\ref{tab:clusters}).  For   {3}  {of such} SCs in the VMR we report a firm detection of emission  {that we can relate to}  stellar winds  {due to the    {superposition} of the high-energy and infrared emission}. Their extracted energy spectra are shown in Fig. \ref{fig:sed}. All the targeted clusters are embedded in a H\textsc{ii} region, clearly identified both at IR, by the WISE-22 $\mu$m survey \cite{Anderson2014TheRegions}, and in optical by observing their characteristic H$\alpha$ emission \cite{Rodgers1960AWay}. 
This further confirms that we are dealing with very massive stars. 
The typical age of embedded clusters is of the order of $\lesssim$ 1 Myr, allowing us to conclude that no SN event happened in these regions yet. The known supernova remnants in the region, Vela Y and Vela Z (also known as Vela Junior) are found in the foreground of the VMR and therefore can be easily masked to avoid any contamination to the $\gamma$-ray signal.
Superposition of other possible accelerators is also excluded as we include in the sample only clusters which are not associated with any identified GeV source from the 4FGL Fermi-LAT (Large Area Telescope) catalog \cite{Abdollahi2022IncrementalCatalog}, nor with any pulsar from the ATNF (Australian Telescope National Facility) catalog  \cite{Manchester2005TheCatalogue}. We note, instead, that at the location of all of the   {considered} SCs, unidentified Fermi-LAT sources are present and we  {suggest an association} with the clusters.
 {Even though no firm identification was proposed so far for these four Fermi-LAT sources, the association we put forward is consistent with tentative classifications reported in earlier works\cite{Malyshev2023Multi-classDefinition}.}
 
The GeV energy band is the optimal range to investigate the global energetic of CR particles, as in that part of the spectrum resides the bulk of their power. 
We analyzed data accumulated during 13 years by the Fermi-LAT in the energy interval between 500 MeV and 1 TeV as described in detail in the Methods section. In this energy range, the LAT point spread function ranges from 1$^\circ$ to 0.15$^\circ$  and its sensitivity for a 10-year long exposure is  $\sim 10^{-12}$ erg  cm$^{-2}$ s$^{-1}$. 
We detected $\gamma$-ray emission from the direction of the SCs: figures~\ref{fig:tsmap} and \ref{fig:sed} show the results of our analysis of the $\gamma$-ray observations carried out in the entire VMR region. Since the targeted objects are low-surface-brightness sources, we tested the stability of the results assuming three different models for the diffuse background emission: the details are reported in the Methods section where we show that the choice of the background model does not affect the detection, nor the spectral shape of the targeted sources, but only slightly (a factor 1.5 at most) the overall flux. We fitted the position and the extension of the sources.     {The results of the morphology tests are presented in Table \ref{tab:hii}.} 
  {In a scenario where stars both heat the dust grains and accelerate CRs we expect a morphological similarity between the gamma-ray and the IR emitting regions. We therefore tested the morphology of the analyzed sources. We found a significant extension ($\sigma_{ext} > 5$) only for the source RCW38, that is found to be comparable with the extension of the HII region. For the others, the extension significance is lower ($\sigma_{ext} \sim 3$), due to the limited angular resolution of Fermi-LAT, but the extension fit still converges to values similar to the extension of the H\textsc{ii} regions (see Table \ref{tab:hii}). In the latter cases, the extension should be regarded as an upper limit to the real extension. The centroid, instead in all cases is found to compare well with the centroid of the corresponding HII regions. The chances for random superposition are low, and are evaluated in the Methods.} 


The most prominent region in $\gamma$ rays corresponds to the location of RCW 38, which is detected with a statistical significance of $\sigma \sim 21$. 
RCW 38 is an ultra-compact H\textsc{ii} region that contains a young ($\sim$~0.5 Myr) stellar cluster with  {more than 1000 stars}, of which at least 30 early type stars are concentrated in a radius of $\sim$ 0.5 pc \cite{Wolk2008HandbookRegions, Wolk2002DISCOVERY38}. 
RCW 38 is embedded in the Vela~B complex, the farthest region within the VMR.  A recent precise determination of its distance has been obtained using Gaia extinction data that located RCW 38 at approximately 1600 pc from us \cite{Zucker2020AHandbook}. The energy spectrum of RCW 38 is modeled as a power law, ${N_0 (E/E_0)^{\alpha}}$   {,with flux normalization $N_0=3.9 \pm 0.2$ MeV$^{-1}$ cm$^{-2}$ s$^{-1}$ at $E_0$=1 GeV and} index $\alpha_\gamma=  {-2.56\pm 0.05}$.   {Its} luminosity above 1 GeV is $L_\gamma (> 1 ~\mathrm{GeV})$=~5~$\times 10^{33}$~erg~s$^{-1}$ , assuming a distance of 1600 pc. 
 The second brightest source in our sample coincides with RCW 36, embedded in the Vela C region, located 900 pc from us \cite{Zucker2020AHandbook}. This region is slightly less luminous (  {$N_0=3.8$ MeV$^{-1}$ cm$^{-2}$ s$^{-1}$ }, $L_\gamma(>1~\mathrm{GeV})$= 1.5 $\times 10^{33}$ erg s$^{-1}$, detected at $\sigma \sim 16$) and  {slightly} steeper ($\alpha_\gamma=   {-2.72 \pm 0.06}$), compared to RCW 38. Similarly to RCW 38, also  {RCW 36}  is ionized by a young ($\sim$ 1 Myr) and compact ($\sim$ 0.5 pc) stellar cluster \cite{Ellerbroek2013Rcw36:Formation}. 
  {Finally, another statistically significant ($\sigma$>5) source is found to be coincident with the young star cluster RCW 32, in the Vela D region, localized at a similar distance than Vela C, $\sim 900$~pc. } 
The age of the RCW 32 is somewhat uncertain because the cluster seems to host two different populations of stars: one made of  very massive stars at the core of the cluster with an age of $\lesssim$ 2 Myr, and a population of low-mass stars that formed earlier (5--13 Myr) \cite{Prisinzano2018AstrophysicsStars}. However, this fact should not cause concern, as older low-mass stars do not end up in a SN explosion. 

Other SCs are located in the VMR (see Table \ref{tab:hii}). We report a hint of $\gamma$-ray emission ($\sigma \gtrsim 5$)  from RCW 35, RCW 37, RCW 40, and RCW 41. Their morphology and spectra have larger uncertainties and therefore we prefer to exclude them from the discussion, {but  {their spectral energy distributions} are reported in the Methods section. }   {Gamma ray emission in proximity of IRS 31 is also detected at a 11 $\sigma$ level. The latter source is not included in the RCW catalog of H$\alpha$ regions \cite{Rodgers1960AWay}  but it is a well known IR source\cite{Massi2003StarC-cloud} coinciding with the WISE region G264.124+01.926.  Most likely, it belongs to the population of radio-quiet HII regions\cite{Anderson2014TheRegions} which, due to their small size are believed to be optically thick to recombination lines (at both optical and radio frequencies), suggesting that also in this case the $\gamma$-ray emission could be connected to massive stars.  {Even if the mass in the region of IRS 31 is quite uncertain (see Table \ref{tab:hii_mass}), this source emerges above the tested backgrounds, but with a reduced significance when changing the background and therefore we ignore it in the discussion.}}
A significant $\gamma$-ray source with extension and position matching RCW 27 is also detected. However, the superposition with a pulsar makes it hard to associate the $\gamma$-ray emission to the embedded stars, and therefore we leave this source for a more detailed investigation.  {No significant emission emerged from RCW 34, this could be due to the vicinity of the source to RCW 36 and IRS 31, but also to its slightly larger distance, $\sim 2.5 \pm 0.2$ kpc \cite{Bik2010SequentialRegions}} 

The detection of these   {three} young SCs in $\gamma$-rays is unique, having no contamination from other sources and a sufficiently precise estimate of their distance and age. This allows us to estimate the fraction of the mechanical power of stellar winds that is converted into CRs.
To estimate the mechanical power of winds in a cluster, $L_*$, we assume that the  {fitted} size of the $\gamma$-ray source corresponds to the radius $R$ of the bubble inflated in the interstellar medium by the SC winds. 
The evolution of the radius of the bubble with the age of the SC, $t$, follows the well known relation\cite{Weaver1977InterstellarEvolution.} $R \sim (\xi L_*/n)^{1/5} t^{3/5}$, where $n$ is the density of the ambient gas, that can be estimated from gas column density measurements performed by making use of tracers such as CO lines and dust emission (see Methods section).
More precisely, the mass determined in this way is assumed to be concentrated within spherical volume of radius equal to $R$, i.e., the measured extension of each $\gamma$-ray source. The depth of the molecular gas along the line-of-sight is certainly larger than $R$, but the dust profiles obtained by Gaia in this region puts an upper limit on the gas extent in Vela to $\sim$ 100 pc \cite{Lallement2022UpdatedDust}. What we obtain in this way is an upper limit on the gas density, that can be converted into an upper limit for the wind mechanical power: $L_* < R^5 t^{-3} n/\xi$.
The parameter $\xi < 1$ accounts for the fact that, due to radiative losses, only a fraction of the mechanical energy $L_*$ is actually used to create the bubble \cite{Yadav2017HowSuperbubbles}. We constrain the value of $\xi$ to match the mechanical luminosity ($\dot{M}v^2$/2) estimated for RCW38, for which a mass-loss rate of $\dot{M}= 2 \times 10^{-4}$ M$_{\odot}$, and an average wind velocity 1000 km s$^{-1}$ are estimated \cite{Canto2000THESTARS}  and find $\xi=0.07$, which is consistent with the values found in hydro-dynamical simulations of very young SCs embedded in a dense ($\sim$ 1000 cm$^{-3}$) medium \cite{Yadav2017HowSuperbubbles}.

The mechanical luminosity obtained in this way is compared to the observed CR luminosity, $L_{CR}$, derived from $\gamma$-ray observations, to compute the acceleration efficiency, $\eta = L_{CR}/L_*$. 
We consider a scenario where the observed $\gamma$-rays are the result of the decay of neutral pions which are in turn generated in the interactions of accelerated CR nuclei with the ambient gas. {We consider leptonic processes due to CR electrons either scattering off the soft ambient photons (inverse Compton scattering) or due to bremsstrahlung less plausible ({see extended discussion in the Methods section}). Synchrotron losses largely dominate in these systems, given the enhanced magnetic field compared to the average in the ISM: for RCW 38 estimations suggest a value of $\sim$ 40 $\mu$G \cite{Bourke2001NEWCLOUDS}, that could be even larger in similar systems\cite{Badmaev2022InsideSimulations}. In such a large magnetic field it is hard to explain the observed spectral points up to 100 GeV with inverse Compton (IC) scattering or bremmstrahlung, as high energy electrons would quickly cool by emitting synchrotron photons in the  radio domain\cite{Padovani2019Non-thermalRegions}. 
On the other hand, the dense gas that constitutes the VMR  {$n\sim$ 1000 cm$^{-3}$ } provides thick target for CR hadronic interactions.

Starting from the observed $\gamma$-ray spectra (Fig.~\ref{fig:sed}), we derived the spectra of the parent CR protons using the \texttt{naima} software package \cite{Zabalza2015}. 
Then, from the derived CR spectral distribution we evaluated the CR luminosity as: $ L_{CR} \approx {W_p}/{t_{pp}}$, where $W_p$ is the total CR proton energy stored within the SC and $t_{pp}$ is the proton-proton interaction timescale \cite{Aharonian2004}. 
Such an estimate of the CR luminosity is based on a calorimetric assumption: all CRs accelerated at the SC lose all their energy due to interactions with the ambient gas over a time which is shorter than both the escape time of CRs from the region and the age of the SC. {  While we know the age ($\sim 10^{13}$ s), the escape time can be computed from the observed $\gamma$-ray and CR luminosity as a function of $\eta$, the fraction of SC mechanical energy that goes into CRs (the derivation is reported in Methods section). For RCW 38 the escape time for 1-GeV particles results $t_{esc}= 7.6 \times 10^{11}  \eta^{-1} ({n}/{1000~ \mathrm{cm^{-3}}})^{-1} ~\mathrm{s} $ which is always smaller than the proton-proton interaction timescale $t_{pp}=1.6 \times 10^{12}  ({n}/{1000~ \mathrm{cm^{-3}}})^{-1} ~\mathrm{s}$ for $\eta>0.005$.  Similarly, the diffusion coefficient  results (see Methods section) $D \sim 2\times 10^{28}~\mathrm{cm^{2} s^{-1}} \eta~(E/1~\mathrm{GeV})^{1/3}({n}/{1000~ \mathrm{cm^{-3}}})$. Thus, even if a strong suppression of $D$ must take place in these systems, compared to the  average Galactic value, $D_{ISM}(1~\mathrm{GeV})\sim 10^{28}~\mathrm{cm^{2} s^{-1}}$, we expect that CRs are efficiently escaping from the systems, meaning that the actual CR luminosity is much higher than the estimations {obtained in the calorimetric hypothesis}. Despite the large target gas density, we don't see $\gamma$-ray emission in the surroundings of the targeted SCs. However, since the SCs themselves have a low surface brightness, on a comparable level to the diffuse emission, it is also possible that escaped particles are too diluted to cause a detectable enhancement. The only hint comes from a gas region near RCW36, approximately at $l,b \sim 266^\circ, 0.9^\circ$, but the uncertainties on its mass prevent us from a firm conclusion on its nature (Figure \ref{fig:tsmap}).}

The CR luminosity calculated in the calorimetric assumption allows us to derive a strict lower limit for $L_{CR}$, as it is based on the most effective conversion of CR energy into $\gamma$-ray photons. It follows that our estimate of $\eta = L_{CR}/L_*$, based on a lower limit for $L_{CR}$ and on a upper limit for $L_*$, is a strict lower limit for the CR acceleration efficiency (see Table~\ref{tab:clusters}).  
 {The values that we obtain are consistent with the upper limits for $\eta$ obtained from the non-detection of a number of embedded star clusters \cite{Maurin2016EmbeddedConstraints}. }

Taking the average value of ${\eta}_{min} \approx 0.5$~\% as representative of all SCs in the Galaxy, stellar winds would accelerate CRs at a rate at least equal to $5 \times 10^{38}$ erg s$^{-1}$,  {corresponding to the 0.7\% of the total energy supply of GCRs}. Remembering that $\eta_{min}$ is a strict lower limit of the CR acceleration efficiency, we conclude that SCs provide a sizable fraction, $\epsilon_w \gg $ 1\%,  of the total power of GCRs. 

 Remarkably, the lower-bound value for  $\epsilon_w = 1$\%  that we obtained from $\gamma$-ray observations implies that an excess in the $^{22}$Ne/$^{20}$Ne ratio in CRs must exist.  {Applying this value, in fact, we find  a lower limit for the isotopic ratio $ X_{CR} \sim \epsilon_w X_w + (1 - \epsilon_w) X_{S} = 0.09 > X_{s}$}, where $X_w \sim 1.56$ is the estimated value for the isotopic ratio in the accelerated stellar wind material\cite{Tatischeff2021TheComposition}. This argument can be reversed to say that, in order to reproduce the observed CR isotopic ratio $X_{CR}\sim 0.317$, the fraction of GCR particles originating from accelerated wind material has to be $\epsilon_w \sim (X_{CR}-X_S)/(X_w-X_S) \sim 16$\%. 
To guarantee that this value is not overshoot, a maximum fraction $\eta_{max} \sim \epsilon_w~L_{tot,CR}/L_{tot,*}\sim 10\%$ of the mechanical energy of the stars may be released in the form of CR nuclei.  Our results indicate that young SCs are non-negligible contributors of galactic CRs and their ability to produce high-energy particles must not be disregarded, especially for their close interconnection to the interstellar medium. 

\begin{table}
\small
\centering
\begin{tabular}{l|cc|cccc|cc}

Source &   $t_{SC}$&  $ n_{ISM}$  {(gas/dust)}&$L^{PL}_\gamma$(>1 GeV) &$N_0$& $\alpha^{PL}_\gamma$ & $R_{\gamma}$ & $L_{CR,min}$(>0.1 GeV)& $\eta_{min} $ \\
			    & [Myr] & [$10^{3}$ cm$^{-3}$] &[10$^{33}$ erg s$^{-1}$]  &[10$^{-12}$ MeV cm$^{-2}$ s$^{-1}$]&   & [$^\circ$] & [10$^{34}$ erg s$^{-1}$]& [\%] \\
\hline
RCW32  & 2 &1.9 {/2.4}     &{0.8}  &1.7 $\pm$ 0.2&  {-2.46 $\pm$ 0.08} & 0.26       &  {0.4}      &  {0.8/0.7}  \\ 
RCW36  & 1.1 &  {2.6/2.9}  &{1.5}    &3.8 $\pm$ 0.2 &-2.72 $\pm$ 0.06 &0.27 &  {1.2}  &   {0.8 /0.7 }  \\
RCW38  &0.5 & 2.1  {/1.9}  &{4.9}    &3.9 $\pm$ 0.2&-2.56 $\pm$ 0.06 & 0.21   &  {3.2}  &   {0.004/0.004 }  \\
\hline
\end{tabular}
\caption{Details on the targeted star clusters: the age of the clusters \cite{Prisinzano2018AstrophysicsStars,Ellerbroek2013Rcw36:Formation,Wolk2002DISCOVERY38,Massi2019DenseC} and density  {(calculated from gas\cite{Dame2000,BenBekhti2016},} and dust\cite{Ade2011} maps) of their surrounding medium are reported along with the measured $\gamma$-ray properties: the $\gamma$-ray extension   {(see also Table \ref{tab:hii})},   {normalization at 1 GeV}, spectral index, and total luminosity, calculated by integrating the best-fit power-law curve above 1 GeV and assuming a common distance of 900 pc for the clusters, except for RCW 38 (1600 pc), and the lower limit for the cosmic-ray luminosity and for the acceleration efficiency  {calculated for both values of density.} }
\label{tab:clusters}
\end{table}

\begin{figure*}
\includegraphics[width=1\linewidth]{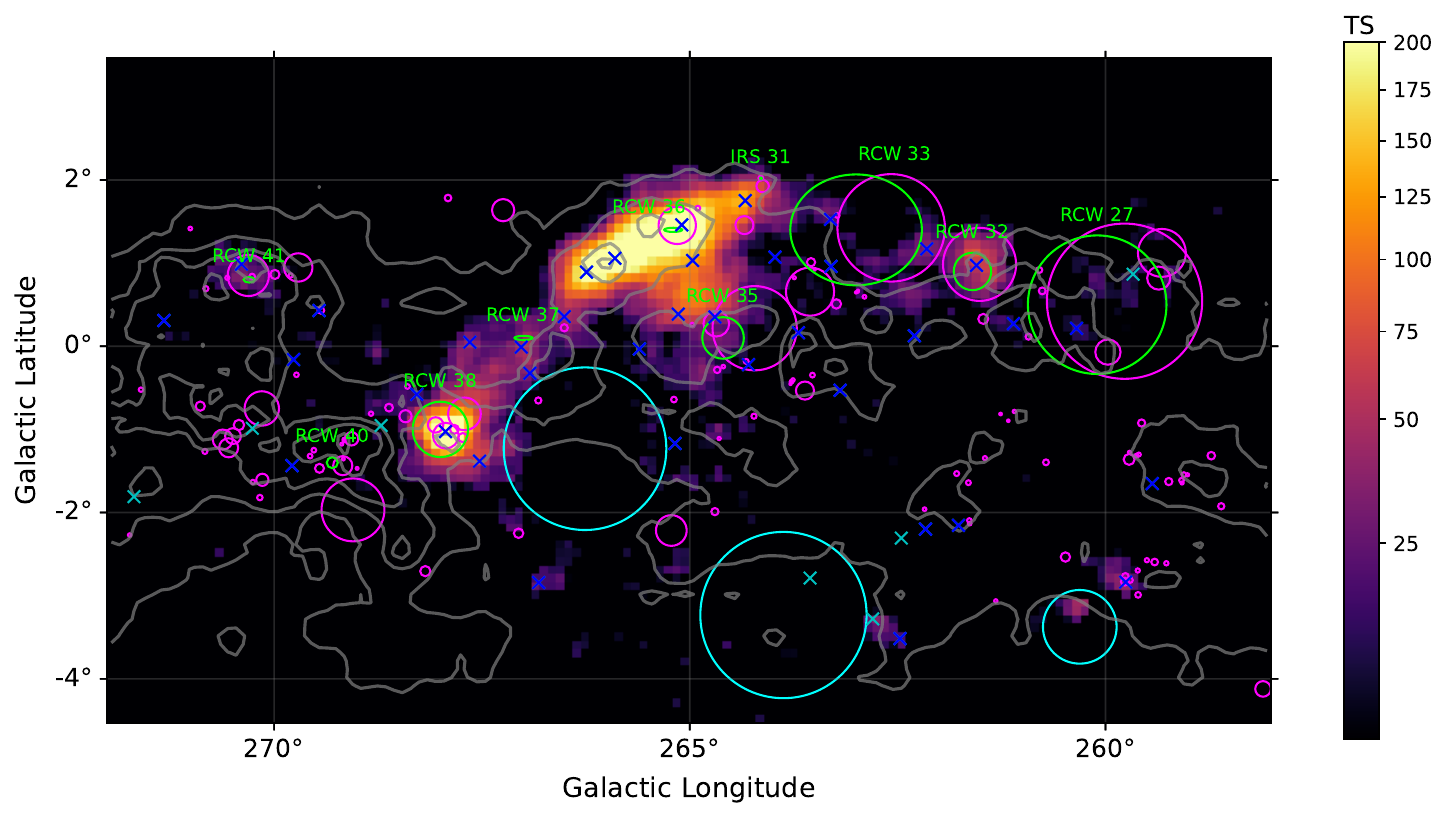}
\caption{{  Test statistic (TS) map  {in the energy range 500 MeV-1 TeV} of the region after the subtraction of the background sources except for the  {sources linked to H\textsc{ii} regions (Table \ref{tab:hii})}. The  {grey} contours represent the gas traced by the (J=$2\rightarrow$1) CO line\cite{Dame2000}; the green and  {magenta} regions are the H\textsc{ii} regions identified by H$\alpha$\cite{Rodgers1960AWay} and IR emission\cite{Anderson2014TheRegions} respectively.  {The blue crosses are the unidentified sources in the ROI, while the cyan crosses and circles are the identified sources of the 4FGL-DR3 catalog \cite{Abdollahi2022IncrementalCatalog}}}.}
\label{fig:tsmap}
\end{figure*}

\begin{figure}
\centering
\includegraphics[width=1 \linewidth]{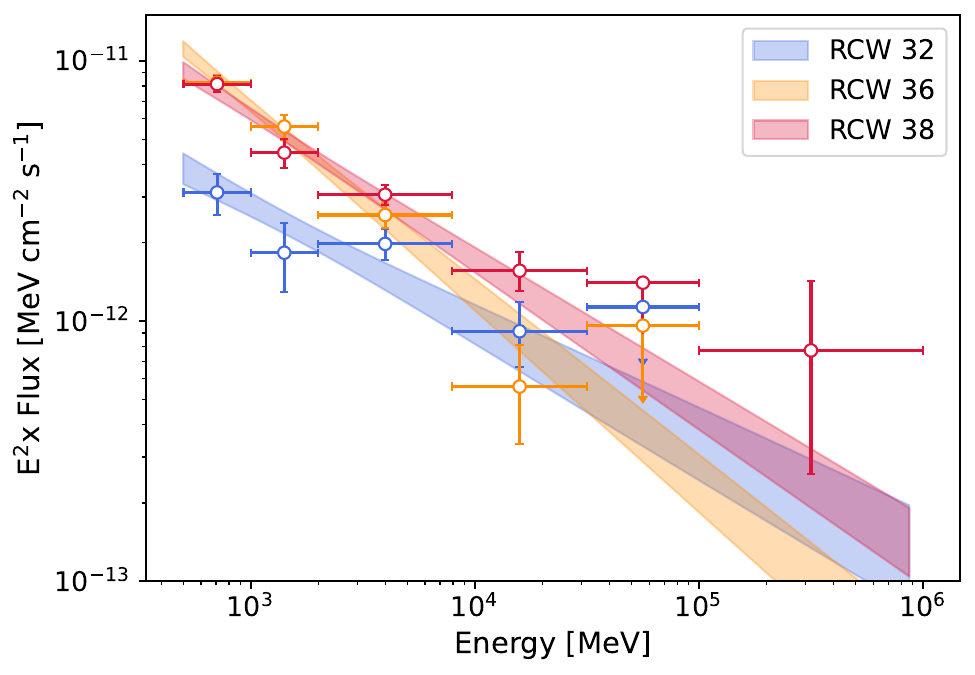}
\caption{Spectral energy distributions of the targeted star clusters.  {The band represents the 1-$\sigma$ uncertainty of the best-fit model. }}
\label{fig:sed}
\end{figure}

\section*{About this manuscript}
This version of the manuscript has been accepted for publication. The accepted manuscript is defined as the version of a manuscript accepted for publication after peer review, but does not reflect post-acceptance improvements, or any corrections, retractions, or other post-publication editorial actions. The Version of Record is available online at: \href{https://www.nature.com/articles/s41550-023-02168-6}{https://www.nature.com/articles/s41550-023-02168-6}.

\section*{Contribution}
G.P lead the data analysis, proposed the interpretation and produced the manuscript and its figures. S.C. proposed the target region as a case study. V.B. cross-checked the Fermi-LAT analysis. G.P, S.G., S.C. and F.A. gave significant inputs on data interpretation. All authors participated in the discussions and editing of the paper. 

\section*{Data availability}
The authors made use of publicly available data that can be retrieved at \href{https://fermi.gsfc.nasa.gov/cgi-bin/ssc/LAT/LATDataQuery.cgi}{https://fermi.gsfc.nasa.gov/cgi-bin/ssc/LAT/LATDataQuery.cgi}

\section*{Code availability}
The authors made use of publicly available analysis softwares. In particular fermipy v.1.0.2 available at \href{https://github.com/fermiPy/fermipy/blob/master/docs/index.rst, naima availble at https://github.com/zblz/naima}{https://github.com/fermiPy/fermipy/blob/master/docs/index.rst, naima availble at https://github.com/zblz/naima}. 

\section*{Competing interest}
The authors declare no competing interests.

\section*{Acknowledgment}
The authors would like to acknowledge Prof. Dr. J. Hinton, Dr. E. Amato, Dr. G. Morlino, Dr. R. Tuffs, and Dr. M. Lemoine-Goumard  for the suggestions and discussion. G. P. and S. G. are supported by Agence Nationale de la Recherche (grant ANR-21-CE31-0028). S. C. acknowledges support from the Polish Science Centre grant DEC-2017/27/B/ST9/02272.

\section*{Methods }\label{methods}

\subsection*{Analysis summary}
We analyzed more than 13 years of Fermi-LAT PASS8 data, collected between August 8th 2009 (MET 239557417) and December 14th 2021 (MET 632287927).   {We selected source class events (\texttt{evclass=128}) reconstructed both at the front and back of the detector  (\texttt{evtype=3}), imposed data quality \texttt{DATA\_QUAL==1 \&\& LAT\_CONFIG==1} and a maximum zenith angle of 90$^{\circ}$, and enabled the energy dispersion evaluation (\texttt{edisp=True}).} 
Due to the large extension of the VMR, we chose a region of interest 17$^\circ$-wide centered at $(l,b) = (265.5,0)^\circ$. In the same region of interest we can find the Vela Junior supernova remnant (4FGL J0851.9-4620e), partially overlapping the cloud complex and the Vela X pulsar (4FGL J0835.3-4510) with its nebula (4FGL J0834.3-4542e). We included in the analysis all the sources from the 4FGL catalog\cite{Abdollahi2022IncrementalCatalog} and the standard galactic (\texttt{gll\_iem\_v07.fits}) and extragalactic (\texttt{iso\_P8R3\_SOURCE\_V3\_v1.txt}) background  provided by the Fermi-LAT collaboration.  { We then tested our detection with different galactic backgrounds based on the map of CO, H\textsc{i} and dust, as explained in the next section.} 

We ran the analysis over the energy range 500 MeV--1 TeV. To investigate the morphology of the sources in the region, we restricted our analysis to data with energy higher than 3 GeV. In this energy range, the Fermi-LAT point spread function ($\sim0.2^\circ$) is much better than at lower energies, where it could be as large as 1$^\circ$.  We deleted from the model all 4FGL sources which spatially coincided with known H\textsc{ii} regions, as reported in Table \ref{tab:hii} and we re-modeled the emission in this regions. In particular, we fitted the position and the extension of these sources  {using the \texttt{extension} routine of the fermipy software package}. Results are collected in Table \ref{tab:hii}:  {we include the best-fit position and extension and their relative uncertainties, along with the $TS_{ext}$ value which gives the significance of the extended source hypothesis over the point-like hypothesis, namely $TS_{ext} = 2\log(L_{ext}/L_{ps} )$, where $L_{ext,ps} $ \cite{Ackermann2017SearchGeV} are the maximum likelihood of the extended and point-like model respectively.  {Although we cannot claim that all the investigated sources are extended, as they are conventionally defined extended if $TS_{ext}>16$ \cite{Ackermann2017SearchGeVb}, we see that the fitted values well agrees with the extension of the corresponding H\textsc{ii} regions. We then used the optimized morphology, obtained at high energies, to extract the spectrum in the whole considered energy range.   {spectral index where the extension hypothesis is indistinguishable from the point-like hypothesis, the fitted extension can be regarded as an upper limit to the real extension.} }
In the table, we also report the significance of the sources from our analysis, determined in the range 500 MeV-1 TeV as $\sigma = \sqrt{-2ln(L_0/L_1)} $ , where $L_1$ and $L_0$ are the value of maximum likelihood obtained for a model with or without the considered source \cite{Mattox1996}. }  {We checked that the total flux and significance did not depend on the uncertainties of the response functions. These can be accounted for by calculating the weighted likelihood \cite{TheFermi-LATcollaboration2019}.     }


\begin{table*}[]
\small
    \centering
    \begin{tabular}{l|llllll}
         Name & WISE name & 4FGL source (type) & Fitted position $(l,b)$& WISE size & Size (TS$_{ext}$) & $\sigma$ \\
         &  & &  [$^{\circ}$] &   [$^{\circ}$]  & [$^{\circ}$] & \\
         \hline 
         RCW 27 & G259.771+00.541 &  J0838.4-3952   (psr) &  259.68$^{\pm   0.10}$,    0.76$^{\pm 0.09}$ & 0.933 & 0.82 \footnote{This value was computed in the energy range 500 MeV-1 TeV, because the source is not significant at higher energeies.}$^{+0.09}_{-0.08}$ ( {46.4}) &   {9.01} \\
         RCW 32 &  G261.515+00.984& J0844.9-4117 (unid)&  261.51$^{\pm 0.06}$,    0.95$^{\pm 0.08}$ & 0.440 & 0.26$^{+0.06}_{- 0.05}$ {(10.7)}  &  {7.48}   \\

         RCW 34 &  G264.343+01.45  & -- & -- &0.108 &-- & --\\
         \multirow{2}{*}{RCW35} & G264.681+00.272 &  \multirow{2}{*}{J0853.1-4407 (unid)}  &  \multirow{2}{*}{264.86$^{\pm 0.08}$,   -0.01$^{\pm 0.12}$} & 0.155 &\multirow{2}{*}{ {n.c.}} & \multirow{2}{*}{ {6.73}} \\
             &  {G264.220+00.216} &  &  & 0.5& & \\
         
         RCW 36 & G265.151+01.454 & J0859.3-4342 (unid)  & 265.09$^{\pm 0.05}$, 1.36$^{\pm 0.04}$ & 0.224 & 0.27$^{+0.07}_{-0.06}$ (10.8) &  {16.14} \\
         RCW 37 & -- &  J0900.2-4608 (unid)& 266.97 $^{\pm   0.04}$,    0.01 $^{\pm 0.05}$ &  -- & point-like &  {4.03} \\
         RCW 38 &  G267.935-01.075 & J0859.2-4729 (unid)& 267.91 $^{\pm0.03 }$,-1.03$^{\pm0.03}$ &  0.155 & 0.21$^{+ 0.04} _{- 0.04}$  {(26.781)} &  {21.04} \\
         RCW 40 & G269.174-01.436 & --&268.57$^{\pm 0.06}$,   -0.73$^{\pm 0.04}$  & 0.115 &  -- & -- \\
         RCW 41 & G270.310+00.851 & J0917.9-4755 (unid)& 270.12 $^{\pm 0.13}$,  0.67$^{\pm 0.11}$ & 0.248 & 0.34$^{+ 0.15}_{- 0.09}$ (6.5) &  {6.96}
 \\
           \hline
         IRS 31 & G264.124+01.926 &  {J0857.7-4256c (unid)}	& 264.28$^{\pm 0.05}$,    1.82 $^{\pm 0.04}$ &  0.075 & 0.14$^{+0.06}_{- 0.05}$  {(4.9)} &  {11.73} \\ 
                 \hline
        \multirow{2}{*}{ {Gas core}} &  \multirow{2}{*}{--} & {J0900.5-4434c (unid)}    &\multirow{2}{*}{ $  {266.13^{\pm 0.05}, 0.86 ^{\pm 0.05}}$} &\multirow{2}{*}{--} & \multirow{2}{*}{ $  { 0.39^{+0.05}_{-0.04} (57)}$} & \multirow{2}{*}{ {21.22}}\\ 
           & &  {J0900.5-4434c (unid)} &  &  &
 &  \\
    \end{tabular}
    \caption{ {Results of the modeling of the considered regions. Along with the results from the extension and position fit, we report the significance, $\sigma$, of the sources computed in the range 500 MeV--1 TeV and the extension significance, $TS_{ext}$, computed above 3 GeV. The instance n.c. indicated those source for which the extension fitting did not converge. }  {The name refers to the} Rodgers-Campbell-Whiteoak (RCW) \cite{Rodgers1960AWay} catalog.  {We report also the name and morphology of the corresponding region found in the WISE catalog \cite{Anderson2014TheRegions} and the name and type of the overlapping Fermi sources from the 12-years source catalog \cite{Abdollahi2022IncrementalCatalog}. }}
    \label{tab:hii}
\end{table*}

\begin{table*}[]
\small
    \centering
    \begin{tabular}{l|llllll}
         Name & Mass\footnote{d=1 kpc} (dust/CO)  & $N$ (dust/CO) & & & & \\
         &  [ 10$^{4}$ M$_\odot$  ] & [10$^{3}$ cm$^{-3}$] \\
         \hline 
         RCW 32 &  0.6 /0.5   & 2.4/1.9 \\
         RCW 36 &  0.5 /0.5   & 2.9/2.6   \\
         RCW 38 & 0.4/0.5  & 1.9/2.1 \\
         IRS 31    & 0.5/0.1 & 7.3/ 2.5 \\
         Gas core  & 3.3/1.5  & 3.4/1.6 \\
           \hline
    \end{tabular}
    \caption{ {Mass of the regions under consideration, calculated from the maps of dust and gas\cite{Ade2011,Dame2000,BenBekhti2016}}   }
    \label{tab:hii_mass}
\end{table*}

\subsubsection*{Systematic uncertainties}
The largest source of uncertainty in the analysis of Fermi-LAT data derives from modeling of the Galactic diffuse emission. We used three different background templates to assess the validity of our results. The Galactic  {diffuse emission} includes the pion decay emission arising from interactions of CR nuclei with the interstellar medium and the inverse Compton (IC) radiation of CR electrons with the interstellar radiation fields. Models for IC emission are usually derived with particles propagation codes, assuming a certain distribution of sources and certain boundary conditions. For what concerns pion emission instead, we tested different models including the latest released galactic background gll\_iem\_v07.fits. In this model the pion emission is based on maps of HI and CO and is adjusted to the data in order to match the dust emission measured by Planck and the measured $\gamma$-ray diffuse emission. The fitting procedure is done independently for different rings at different galacto-centric distances, in order to account for possible variation in the CR density towards the Galactic center. For the IC component, the standard background provided by the Fermi collaboration uses a model computed with the \texttt{galprop} propagation code \cite{Vladimirov2011}. We do the same  {for our customized background}, choosing the output configuration \texttt{$^\textrm{S}$Y$^\mathrm{Z}$10$^\mathrm{R}$30$^\mathrm{T}$150$^\mathrm{C}$2}, which assumes the source distribution from Yusifov and Kucuk\cite{Yusifov2004}, and the boundary conditions on the scale of the galactic disk: R=30 kpc, h=10 kpc. We constructed, instead, the model for the pion emission  by using spatial maps of the gas derived by CO\cite{Dame2000} and H\textsc{i} \cite{BenBekhti2016} line emission, and by dust opacity at 353 Hz \cite{Ade2011}. The spectrum is modeled as a broken power law and is fitted to the data. Differently from the standard background, a single spectrum is assumed here for the entire line of sight. This is justified, since the gas in the region of Vela is all concentrated around the distance of $\sim$ 1 kpc, with the only exception of the region of Vela B which is located at $\sim$ 2 kpc. This means that the entire molecular cloud complex is concentrated in $\lesssim$ 1 kpc and therefore one should not expect a significant variation in the CR spectrum here. Spectra obtained with the different backgrounds are reported in Fig. \ref{fig:sys}. The detection of the sources and the general trend is confirmed with all the backgrounds with the exception of the region corresponding to the peak of gas, which is detected only with the standard and CO based background, but disappears when using a dust template.   {The values of flux calculated in the different cases vary by a factor 0.2-0.5 when using a gas template and of 0.3-0.8 when using dust. As a further check, we compared the results of our analysis to the values reported in the 4FGL-DR3 catalog \cite{Abdollahi2022IncrementalCatalog} and saw that the resulting fluxes differ at most by 20\% compared to the tabulated values.   {Meanwhile, the new version of the catalog (DR4) has been announced  and is available in a preliminary version\cite{Ballet2023}. We checked and confirm that all sources discussed are still detected. While the significance of the sources slightly changes from one to another catalog, probably due to different modeling of the background sources, their flux is confirmed within uncertainties.} Other instrumental systematic uncertainties are much smaller, and therefore can be disregarded.  }

\subsection*{Emissivity}
We computed (Table \ref{tab:hii_mass})  the mass from the observed column density of gas and dust using the conversion factors $X_{CO} = 2 \times 10^{20}$ cm$^{-2}$ K$^{-1}$ km$^{-1}$ s \cite{Bolatto2013}  and $X_{dust} = (\tau_D/N_H)^{-1} = 8.3 \times 10^{25}$ cm$^{-2}$ ~\cite{Ade2011}. Notice that in general the amount of gas and dust in each region is consistent, apart from the gas-core region  {and IRS 31}, which  {are} 3-times more massive in the dust template. In principle, this could suggest a component of dark gas, i.e. a region where gas tracers are saturated and therefore fail to account for all the gas. To  {understand if this gamma-ray excess in correspondence of the gas core could be due to a mis-calculation of the mass of the gas,} we extracted the spectrum of this region and compared it to the spectrum of the large-scale region associated to the VMR.  {We used the customized template (based on HI and CO) and isolated from the background the region of the VMR, selected with the condition that the column density exceeded 10$^{22}$ cm$^{-2}$. This region has a total mass of 2.8$\times 10^6$ M$_\odot$.  } 
  {The resulting emissivity, defined as the flux per hydrogen atom, is shown in Fig. \ref{fig:dark}  together with the emissivity derived from diffuse emission in the region within 8 and 10 kpc from the Galactic center \cite{Acero2016}. The gas region of the VMR perfectly matches the emissivity of the diffuse emission, with a spectral index of $  {-2.557\pm -0.003}$. } For the gas core instead, the spectrum is much harder ($\alpha_\gamma =-2.4   {\pm 0.05}$) which suggests a different origin for the emission than untraced gas. We compared also the emission of the H\textsc{ii} regions with the spectrum of the CR sea. The spectra, normalized to the mass of the hydrogen in that regions are shown in Fig \ref{fig:dark}.  {These need to be interpreted as the level of excess of these sources over the diffuse emissivity. This is because the underlying VMR is part of the background and is therefore subtracted from the flux of these regions. } We see a significant variation of the spectra compared to the one of the diffuse gas. To explain the radiation in terms of a dark component, a localized enhancement of the gas density would be needed as the enhancement varies case by case from 0.5 to 3 times (see Fig.\ref{fig:dark}).

\begin{figure}
\includegraphics[width=1 \linewidth]{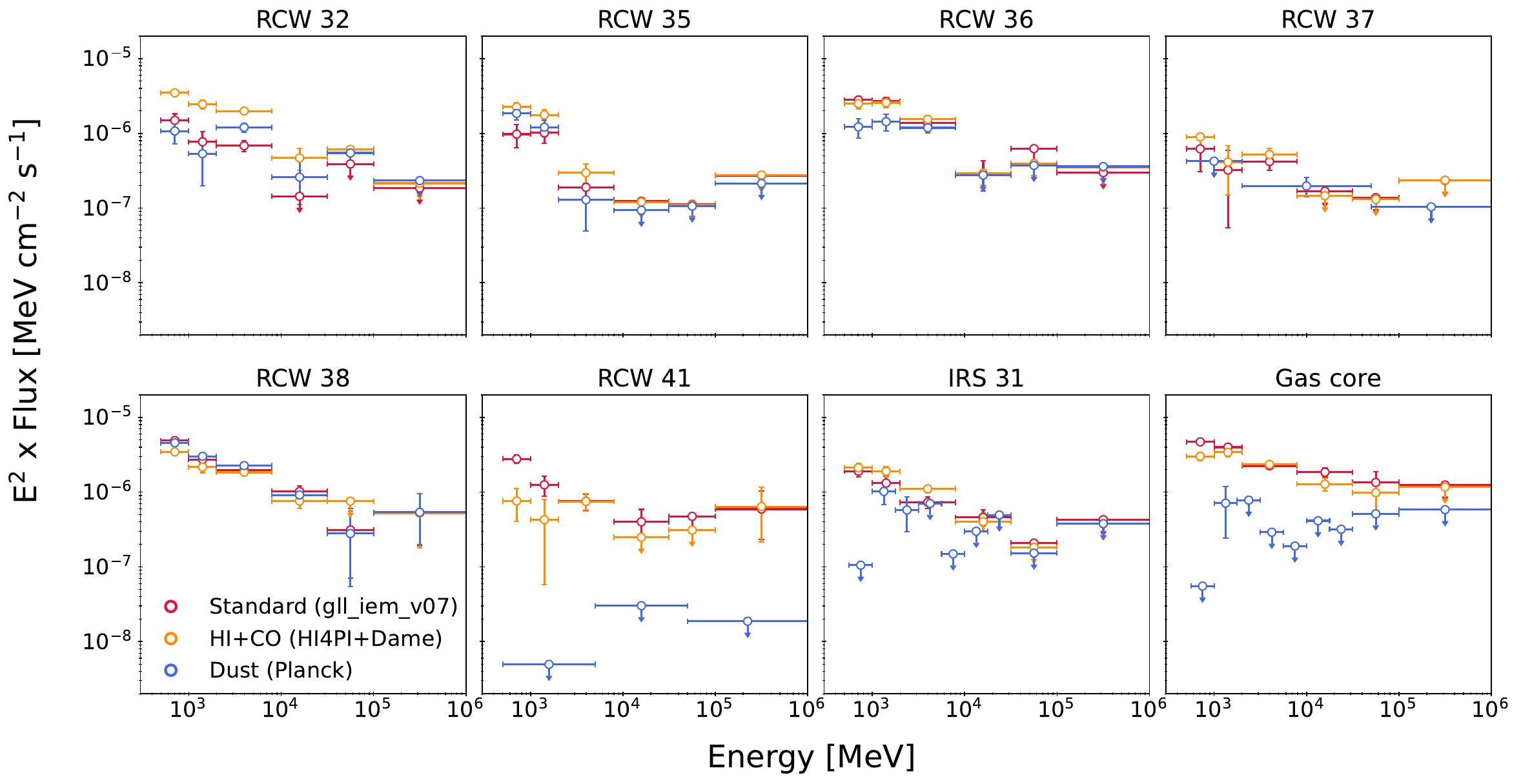}
\caption{Spectra of all sources, obtained with different backgrounds:  {the standard galactic background provided by the Fermi collaboration  {(red points)} and two customized backgrounds based on dust  {(blue points)} and gas  {  {(yellow points)}} as a template for the pion emission.}}
\label{fig:sys}
\end{figure}

\begin{figure}
\includegraphics[width=1 \linewidth]{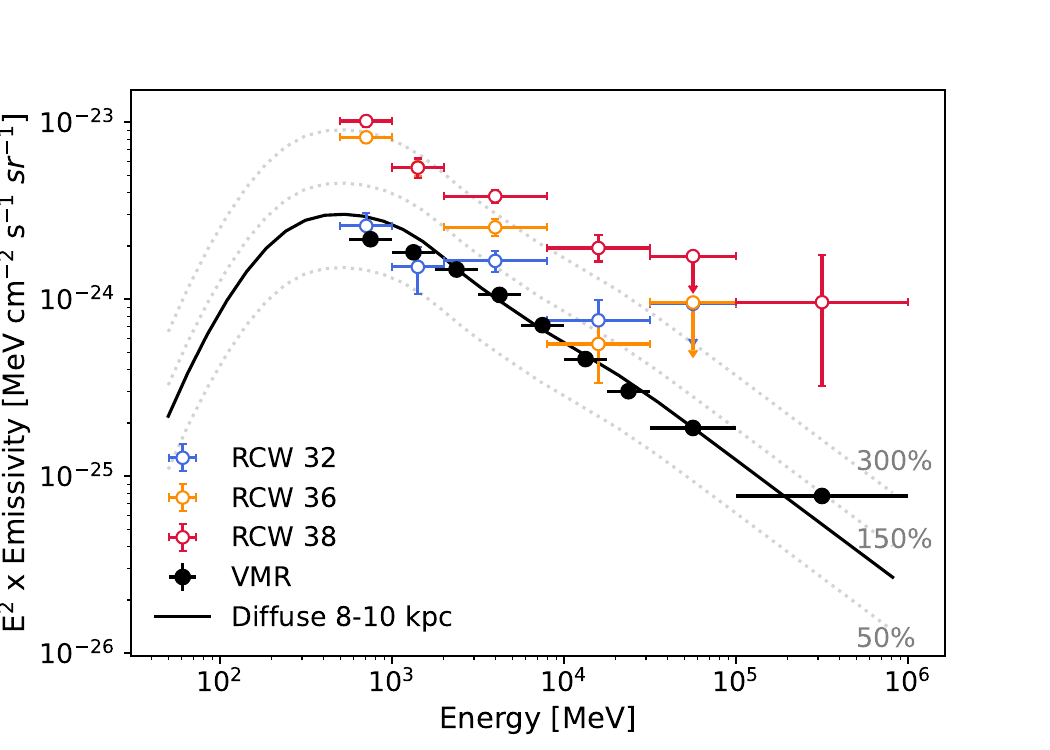}
\caption{ {Excess} emissivity of the H\textsc{ii} sources (normalized by the average mass of the gas in the regions) compared to the emissivity per H atoms of the entire VMR. The used masses are reported in Table \ref{tab:hii_mass}.  The solid curves represent the local $\gamma$-ray emissivity (black) as measured by Fermi-LAT in the 8--10 kpc ring \cite{Acero2016}.  {The dotted lines indicates different levels of excess (from 10\% to 300\%) over the local emissivity. Notice that the errorbars are re-scaled as the flux by dividing by the total number of protons. } }
\label{fig:dark}
\end{figure}

\subsection*{Chance-correlation evaluation}
 {Previous attempt of understanding the nature of the unidentified regions that we targeted were inconclusive. However, in a recent analysis Malyshev and collaborators\cite{Malyshev2023Multi-classDefinition}  investigated the 4FGL unidentified sources by studying their possible association with typical source classes. They created 6 classes each including different types of sources and, using random forest (RF) and neural networks (NN) algorithms, they calculate the probability of each unidentified source of belonging to a certain class. Remarkably the sources we investigate have a large probability of belonging to class 6, that includes galactic sources, namely: binaries, pulsars, star-forming regions, supernova remnants and pulsar wind nebulae. In all the cases considered for the discussion the probability of belonging to this class exceeded 16\%, in the two brightest cases the probability was larger than 40\%. This argument supports our identification with stellar sources, although it should be noticed that these type of classification based on the spectral shape can not predict a new source class, and that embedded star clusters, as the ones we target could have very different spectral characteristic compared to other identified star-forming regions. }

 {In the special case of the HII regions in Vela however, the morphological similarity between the gamma and the IR emission is an additional argument in favor of this identification. We evaluated if the superposition between Fermi and WISE sources in the region of Vela could be a chance coincidence. The 4FGL-DR3 catalog contains 2082 sources with no association. In the RoI, they match with 10 sources from the WISE catalog. We consider it a match if the Fermi source centroid falls within the radius of the HII region. We then simulated 1000 synthetic populations of 2082 sources, by performing Monte Carlo extraction over the longitude and latitude distribution of the unidentified Fermi sources. We repeated the matching procedure with the simulated  sources, the distribution of the number of matches is plotted in Figure \ref{fig:random}. We notice that with the simulations we never obtain a value as high as 10, meaning that the chances for random correlation are smaller than 0.1\%. The average of the matches with the simulated populations is instead 0.4. }
\begin{figure}
    \centering
    \includegraphics[width=1\linewidth]{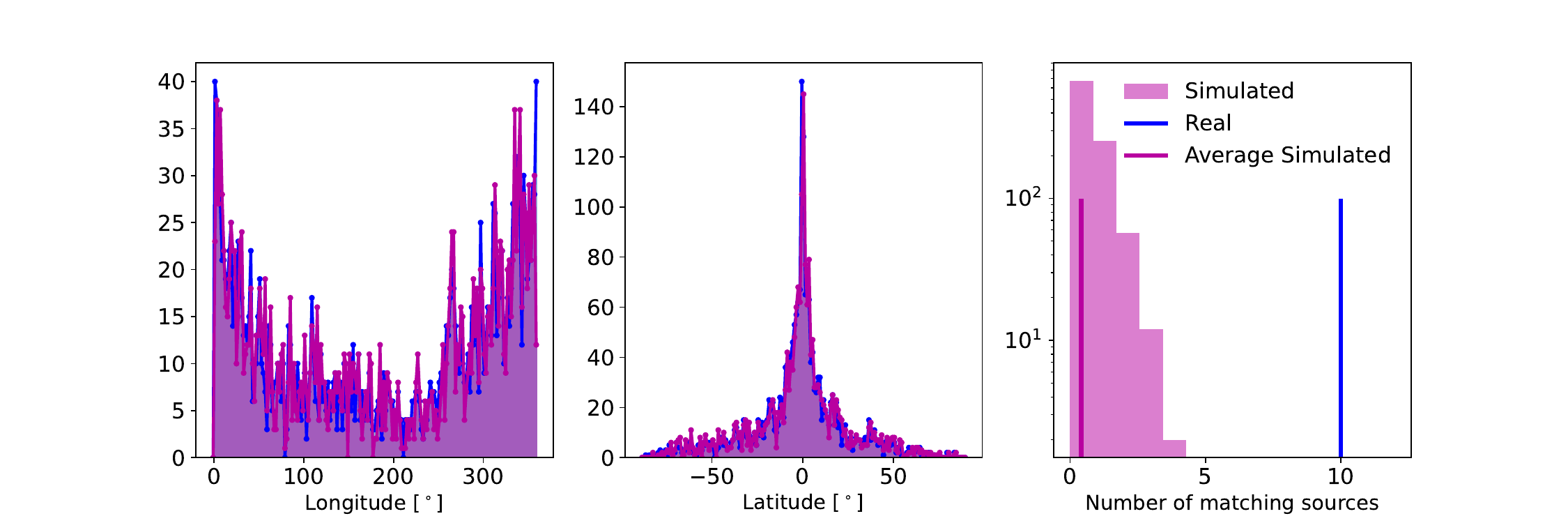}
    \caption{The left and central shows the longitude and latitude distribution of the Fermi unidentified sources and of one example of the simulated sample. In the right panel the number of matches with HII regions in the RoI is displayed in the case of simulated and real Fermi unidentified sources.  }

    \label{fig:random}
\end{figure}

\subsection*{Leptonic scenarios}
 {The detected sources are located in an environment with a large target gas density  ($\sim 10^3$ cm$^{-3}$) and high magnetic field ($B\sim 40~\mu$G ), therefore nuclear interactions are naturally expected to dominate the emission, having a shorter timescale and being the electrons suppressed by cooling. One can see that in these conditions the observed spectra cannot be associated with a leptonic scenario. In the case of inverse Compton scattering on a photon field of energy $\epsilon$, for a given injection rate of electrons, $q(t)=q_0E^{-\alpha_{e}}$, the observed gamma-ray spectrum would be: }
$$Q_{\gamma}(E_{\gamma}) \sim q(E)\tau_{cool} \bigg(\frac{dE}{dt}\bigg)_{IC} \frac{dE}{dE_{\gamma} }\frac{1}{E_{\gamma}}  $$
 {We assume as a characteristic timescale $\tau_{cool}= \frac{E}{(\frac{dE}{dt})_{IC+SYN}}$  the cooling timescale due to IC and synchrotoron interactions. The latter is proportional {to the radiation and magnetic energy density respectively:} $\big(\frac{dE}{dt}\big)_{IC+SYN} = A \omega_{R/B}.$  {The fraction of energy transferred from electrons to $\gamma$-rays is:} $E_{\gamma} = \frac{4}{3} \bigg( \frac{E}{mc^2}\bigg)^{2} \epsilon \propto E^2$, therefore:}
$$Q_{\gamma} = Q_{0,\gamma}E^{-\alpha_\gamma} =q_0 ~E~\frac{A\omega_{R}}{A\omega_R + A \omega_B} \frac{dE}{dE_{\gamma}}\frac{1}{E_{\gamma}} \propto E^{-(\frac{\alpha_e}{2}+1)}_{\gamma} .$$
 {This implies that, in order to obtain the observed gamma-ray spectrum $\alpha_\gamma\gtrsim$ 2.6, an injection spectral index of $\alpha_e \gtrsim 3.2$ is required, in  the assumption where electrons cool via synchrotron and inverse Compton radiation. The timescale for IC cooling depends on energy and on the radiation field energy density. We estimated the radiation density in the NIR, by integrating the maps provided by WISE at 22~$\mu$m regions and we found values spanning from 0.5 to 50~eV~cm$^{-3}$. Assuming the latter as nominal value, we obtain a cooling time of  $6\times 10^7 \mathrm{~yr} (E/1~\mathrm{GeV})^{-1}$, which would exceed the age of the SCs.
In the absence of cooling, the injection spectrum should be even softer to justify the observed emission. In this case, in fact, $\tau=\tau_{age}$ and one obtains $\alpha_\gamma=(\alpha_e+1)/2$ which translates to $\alpha_e=4.2$ for $\alpha_\gamma=2.6$ .}

 {On the other hand bremsstrahlung in a dense environment is as fast as proton-proton interaction, $\sim 5\times 10^5 \mathrm{~yr}$, for densities of 1000 cm$^{-3}$ , but in order to explain the emission via bremsstrahlung, a large electron/proton rate would be required. We can roughly estimate the ratio of electron over proton needed to explain the observed gamma-ray luminosity in terms of bremsstrahlung rather than protons. In the bremsstrahlung/hadronic dominated phase we would have $L_{\gamma}= W_{e}/\tau_{bremss} $,   and in the hadronic-dominated scenario we would have $L_{\gamma}= W_p/\tau_{pp}$ , where $$ W_e=A_{e}\int_{E_{e,min}}^{E_{e,max} }E~E^{-\alpha_e} $$ and $$ W_p=A_{p}\int_{E_{p,min}}^{E_{p,max} }E~E^{-\alpha_p} $$
are the total energy in electron and protons respectively. By equating the two quantities and assuming that in the observed energy range $E_{p,min/max}=10~E_{e,min/max} $ and that the gamma-ray distribution follows the distribution of the parent particles: $\alpha_{e}=\alpha_{p}=\alpha_{\gamma}$ one can obtain $K_{ep}=A_e/A_p$. With our spectral index we obtain values from 0.05 to 0.25, which is much larger that what is estimated in the contest of DSA theory, namely $10^{-2}-10^{-3}$ ~\cite{Park2015SimultaneousShocks}
}
\subsection*{Cosmic ray derivation }
\texttt{Naima} \cite{Zabalza2015} is a software package that allows the derivation of the parent CR distribution given a certain $\gamma$-ray SED. Here we extracted the proton spectrum from the SEDs of the SCs seen by Fermi-LAT  {and shown in Fig. \ref{fig:sed}}, assuming that pion decay is the main radiation mechanism. By default, the software uses the parametrization of the proton-proton cross section derived by Kafexhiu and collaborators \cite{Kafexhiu2014}. We assumed a Power-law spectrum for the protons' total energy, $E$:
$$\frac{dN}{dE} = N_0 \bigg(\frac{E}{E_0} \bigg)^{-\alpha_{CR}} $$
The fitted parameters are reported in Table \ref{tab:naima}. The total kinetic energy of protons, $W_p$, is also reported. The latter is calculated as $\int dT~ T~ \frac{dN}{dT}$, remembering that $E = T + m_p c^2$.

\begin{table}
\centering
\begin{tabular}{l|ccc}
 & $N_{0}$ & $\alpha_{CR}$ & $W_p$ \\ 
 & [10$^{35}$ eV$^{-1}$] & & [10$^{46}$ erg] \\
 \hline
RCW 32 &  {2.33$^{+0.2}_{-0.3}$} &  { -2.5$^{+0.07}_{-0.05}$ }  &  {1.3} \\
RCW 36 & 2.8$^{+0.4}_{-0.3}$  & -2.86$^{+0.08}_{-0.06}$  & 0.63\\
RCW 38 & 4.3$^{+0.3}_{-0.3}$  & -2.71$^{+0.07}_{-0.05}$  & 2.71 \\
\hline

\end{tabular}
\caption{Derived parameters for proton spectra. $N_0$ is calculated for $E_0=$ 50 GeV}
\label{tab:naima}
\end{table}

\subsection*{Diffusion timescale}
We compute the diffusion coefficient from the derived proton energy distribution $\frac{dN}{dE}$ derived with naima. In a situation where particle escape the system, the escape time shapes the observed distribution as: $$\frac{dN}{dE}= q(E)\tau_{esc} =q(E) \frac{R^2}{D_0 (\frac{E}{E_0})^\delta}. $$ This expression can be used to compute the CR luminosity, $L_{CR}=\eta L_{mecc}$, which we assume to be a fraction $\eta$ of the mechanical luminosity: $$ L_{CR}=\eta L_{mecc} = \int dE E 
\frac{dN}{dE}$$ and therefore derive $\tau_{esc}$ and $D_0$ as a function of $\eta$. We compute this for RCW 38, using its mechanical luminosity which is constrained both by  wind-bubble theory\cite{Weaver1977InterstellarEvolution.} and by simulations. We find for this system $D_0 = \eta ~ 1.7 \times 10^{28}$ cm$^{2}$ s$^{-1}$ and $\tau_{esc}= \eta^{-1}~ 7.6 \times 10^{11}$ s.

\bibliography{bib}

\begin{thebibliography}{10}
\urlstyle{rm}
\expandafter\ifx\csname url\endcsname\relax
  \def\url#1{\texttt{#1}}\fi
\expandafter\ifx\csname urlprefix\endcsname\relax\def\urlprefix{URL }\fi
\expandafter\ifx\csname doiprefix\endcsname\relax\def\doiprefix{DOI: }\fi
\providecommand{\bibinfo}[2]{#2}
\providecommand{\eprint}[2][]{\url{#2}}

\bibitem{Seo2018TheProduction}
\bibinfo{author}{Seo, J.}, \bibinfo{author}{Kang, H.} \& \bibinfo{author}{Ryu,
  D.}
\newblock \bibinfo{journal}{\bibinfo{title}{{The contribution of stellar winds
  to cosmic ray production}}}.
\newblock {\emph{\JournalTitle{Journal of the Korean Astronomical Society}}}
  \textbf{\bibinfo{volume}{51}}, \bibinfo{pages}{37--48},
  \doiprefix\url{10.5303/JKAS.2018.51.2.37} (\bibinfo{year}{2018}).

\bibitem{Strong2010GlobalWay}
\bibinfo{author}{Strong, A.~W.} \emph{et~al.}
\newblock \bibinfo{journal}{\bibinfo{title}{{Global cosmic-ray related
  luminosity and energy budget of the Milky Way}}}.
\newblock {\emph{\JournalTitle{Astrophys.J.Lett.}}}
  \textbf{\bibinfo{volume}{722}}, \bibinfo{pages}{L58--L63},
  \doiprefix\url{10.1088/2041-8205/722/1/L58} (\bibinfo{year}{2010}).

\bibitem{Cesarsky1983a}
\bibinfo{author}{Cesarsky, C.~J.} \& \bibinfo{author}{Montmerle, T.}
\newblock \bibinfo{journal}{\bibinfo{title}{{Gamma rays from active regions in
  the galaxy: The possible contribution of stellar winds}}}.
\newblock {\emph{\JournalTitle{Space Science Reviews}}}
  \textbf{\bibinfo{volume}{36}}, \bibinfo{pages}{173--193},
  \doiprefix\url{10.1007/BF00167503} (\bibinfo{year}{1983}).

\bibitem{Casse1982OnRays}
\bibinfo{author}{Casse, M.} \& \bibinfo{author}{Paul, J.~A.}
\newblock \bibinfo{journal}{\bibinfo{title}{{On the stellar origin of the Ne-22
  excess in cosmic rays}}}.
\newblock {\emph{\JournalTitle{The Astrophysical Journal}}}
  \textbf{\bibinfo{volume}{258}}, \doiprefix\url{10.1086/160132}
  (\bibinfo{year}{1982}).

\bibitem{Binns2005Cosmic-RayRays}
\bibinfo{author}{Binns, W.~R.} \emph{et~al.}
\newblock \bibinfo{journal}{\bibinfo{title}{{Cosmic-Ray Neon, Wolf-Rayet Stars,
  and the superbubble origin of Galactic Cosmic Rays}}}.
\newblock {\emph{\JournalTitle{The Astrophysical Journal}}}
  \textbf{\bibinfo{volume}{634}}, \doiprefix\url{DOI 10.1086/496959}
  (\bibinfo{year}{2005}).

\bibitem{Boschini2020InferenceFramework}
\bibinfo{author}{{Boschini}, M.~J.} \emph{et~al.}
\newblock \bibinfo{journal}{\bibinfo{title}{{Inference of the Local
  Interstellar Spectra of Cosmic-Ray Nuclei Z {\ensuremath{\leq}} 28 with the
  GALPROP-HELMOD Framework}}}.
\newblock {\emph{\JournalTitle{apjs}}} \textbf{\bibinfo{volume}{250}},
  \bibinfo{pages}{27}, \doiprefix\url{10.3847/1538-4365/aba901}
  (\bibinfo{year}{2020}).
\newblock \eprint{2006.01337}.

\bibitem{Aharonian2019}
\bibinfo{author}{Aharonian, F.}, \bibinfo{author}{Yang, R.} \&
  \bibinfo{author}{de~O{\~{n}}a~Wilhelmi, E.}
\newblock \bibinfo{journal}{\bibinfo{title}{{Massive stars as major factories
  of Galactic cosmic rays}}}.
\newblock {\emph{\JournalTitle{Nature Astronomy}}}
  \textbf{\bibinfo{volume}{3}}, \bibinfo{pages}{561--567},
  \doiprefix\url{10.1038/s41550-019-0724-0} (\bibinfo{year}{2019}).

\bibitem{Abeysekara2021}
\bibinfo{author}{Abeysekara, A.~U.} \emph{et~al.}
\newblock \bibinfo{journal}{\bibinfo{title}{{HAWC observations of the
  acceleration of very-high-1energy cosmic rays in the Cygnus Cocoon}}}.
\newblock {\emph{\JournalTitle{Nature Astronomy}}} \bibinfo{pages}{1--7},
  \doiprefix\url{10.1038/s41550-021-01318-y} (\bibinfo{year}{2021}).

\bibitem{Cao2021}
\bibinfo{author}{Cao, Z.} \emph{et~al.}
\newblock \bibinfo{journal}{\bibinfo{title}{{Ultrahigh-energy photons up to 1.4
  petaelectronvolts from 12 {$\gamma$}-ray Galactic sources}}}.
\newblock {\emph{\JournalTitle{Nature}}} \textbf{\bibinfo{volume}{594}},
  \bibinfo{pages}{33--36}, \doiprefix\url{10.1038/s41586-021-03498-z}
  (\bibinfo{year}{2021}).

\bibitem{Higdon2005OBRays}
\bibinfo{author}{Higdon, J.~C.}, \bibinfo{author}{Lingenfelter, R.~E.},
  \bibinfo{author}{Higdon, J.~C.} \& \bibinfo{author}{Lingenfelter, R.~E.}
\newblock \bibinfo{journal}{\bibinfo{title}{{OB Associations,
  Supernova-generated Superbubbles, and the Source of Cosmic Rays}}}.
\newblock {\emph{\JournalTitle{ApJ}}} \textbf{\bibinfo{volume}{628}},
  \bibinfo{pages}{738--749}, \doiprefix\url{10.1086/430814}
  (\bibinfo{year}{2005}).

\bibitem{Bykov2015NonthermalSuperbubbles}
\bibinfo{author}{Bykov, A.}
\newblock \bibinfo{journal}{\bibinfo{title}{{Nonthermal particles and photons
  in starburst regions and superbubbles}}}.
\newblock {\emph{\JournalTitle{The Astronomy and Astrophysics Review}}}
  \textbf{\bibinfo{volume}{22}}, \bibinfo{pages}{77},
  \doiprefix\url{10.1007/s00159-014-0077-8} (\bibinfo{year}{2015}).

\bibitem{Morlino2021ParticleClusters}
\bibinfo{author}{Morlino, G.}, \bibinfo{author}{Blasi, P.},
  \bibinfo{author}{Peretti, E.} \& \bibinfo{author}{Cristofari, P.}
\newblock \bibinfo{journal}{\bibinfo{title}{{Particle acceleration in winds of
  star clusters}}}.
\newblock {\emph{\JournalTitle{Monthly Notices of the Royal Astronomical
  Society}}} \textbf{\bibinfo{volume}{504}}, \bibinfo{pages}{6096--6105},
  \doiprefix\url{10.1093/mnras/stab690} (\bibinfo{year}{2021}).

\bibitem{Vieu2022CosmicSuperbubbles}
\bibinfo{author}{Vieu, T.}, \bibinfo{author}{Gabici, S.},
  \bibinfo{author}{Tatischeff, V.} \& \bibinfo{author}{Ravikularaman, S.}
\newblock \bibinfo{journal}{\bibinfo{title}{{Cosmic ray production in
  superbubbles}}}.
\newblock {\emph{\JournalTitle{Monthly Notices of the Royal Astronomical
  Society}}} \textbf{\bibinfo{volume}{512}}, \bibinfo{pages}{1275--1293},
  \doiprefix\url{10.1093/mnras/stac543} (\bibinfo{year}{2022}).

\bibitem{Aharonian2022A1}
\bibinfo{author}{Aharonian, F.} \emph{et~al.}
\newblock \bibinfo{journal}{\bibinfo{title}{{A deep spectromorphological study
  of the {$\gamma$} -ray emission surrounding the young massive stellar cluster
  Westerlund 1}}}.
\newblock {\emph{\JournalTitle{Astronomy and Astrophysics}}}
  \textbf{\bibinfo{volume}{666}}, \bibinfo{pages}{A124},
  \doiprefix\url{10.1051/0004-6361/202244323} (\bibinfo{year}{2022}).

\bibitem{Wright2015TheOB2}
\bibinfo{author}{Wright, N.~J.}, \bibinfo{author}{Drew, J.~E.} \&
  \bibinfo{author}{Mohr-Smith, M.}
\newblock \bibinfo{journal}{\bibinfo{title}{{The massive star population of
  Cygnus OB2}}}.
\newblock {\emph{\JournalTitle{Monthly Notices of the Royal Astronomical
  Society}}} \textbf{\bibinfo{volume}{449}},
  \doiprefix\url{10.1093/mnras/stv323} (\bibinfo{year}{2015}).

\bibitem{Mestre2021Probing2}
\bibinfo{author}{Mestre, E.} \emph{et~al.}
\newblock \bibinfo{journal}{\bibinfo{title}{{Probing the hadronic nature of the
  gamma-ray emission associated with Westerlund 2}}}.
\newblock {\emph{\JournalTitle{Monthly Notices of the Royal Astronomical
  Society}}} \textbf{\bibinfo{volume}{505}},
  \doiprefix\url{10.1093/mnras/stab1455} (\bibinfo{year}{2021}).

\bibitem{Maurin2016EmbeddedConstraints}
\bibinfo{author}{Maurin, G.}, \bibinfo{author}{Marcowith, A.},
  \bibinfo{author}{Komin, N.}, \bibinfo{author}{Krayzel, F.} \&
  \bibinfo{author}{Lamanna, G.}
\newblock \bibinfo{journal}{\bibinfo{title}{{Embedded star clusters as sources
  of high-energy cosmic rays Modelling and constraints}}}.
\newblock {\emph{\JournalTitle{Astronomy {\&} Astrophysics}}}
  \textbf{\bibinfo{volume}{591}}, \doiprefix\url{10.1051/0004-6361/201628465}
  (\bibinfo{year}{2016}).

\bibitem{Ekstrom2012Grids0.014}
\bibinfo{author}{Ekstr{\"{o}}m, S.} \emph{et~al.}
\newblock \bibinfo{journal}{\bibinfo{title}{{Grids of stellar models with
  rotation - I. Models from 0.8 to 120 Msun at solar metallicity (Z =
  0.014)}}}.
\newblock {\emph{\JournalTitle{Astronomy {\&} Astrophysics}}}
  \textbf{\bibinfo{volume}{537}}, \bibinfo{pages}{A146},
  \doiprefix\url{10.1051/0004-6361/201117751} (\bibinfo{year}{2012}).

\bibitem{Wood1989MASSIVEGALAXY}
\bibinfo{author}{Wood, D. O.~S.}, \bibinfo{author}{Churchwell, E.} \&
  \bibinfo{author}{Observatory, W.}
\newblock \bibinfo{journal}{\bibinfo{title}{{Massive Stars Embedded in
  Molecular Clouds: Their Population and Distribution in the Galaxy}}}.
\newblock {\emph{\JournalTitle{The Astrophysical Journal}}}
  \textbf{\bibinfo{volume}{340}}, \bibinfo{pages}{265--272},
  \doiprefix\url{10.1086/167390} (\bibinfo{year}{1989}).

\bibitem{Mascoop2021TheWavelengths}
\bibinfo{author}{Mascoop, J.~L.} \emph{et~al.}
\newblock \bibinfo{journal}{\bibinfo{title}{{The Galactic H ii Region
  Luminosity Function at Radio and Infrared Wavelengths}}}.
\newblock {\emph{\JournalTitle{The Astrophysical Journal}}}
  \textbf{\bibinfo{volume}{910}}, \bibinfo{pages}{159},
  \doiprefix\url{10.3847/1538-4357/abe532} (\bibinfo{year}{2021}).

\bibitem{Anderson2014TheRegions}
\bibinfo{author}{Anderson, L.~D.} \emph{et~al.}
\newblock \bibinfo{journal}{\bibinfo{title}{{The wise catalog of galactic HII
  regions}}}.
\newblock {\emph{\JournalTitle{Astrophysical Journal, Supplement Series}}}
  \textbf{\bibinfo{volume}{212}}, \bibinfo{pages}{1},
  \doiprefix\url{10.1088/0067-0049/212/1/1} (\bibinfo{year}{2014}).

\bibitem{Rodgers1960AWay}
\bibinfo{author}{Rodgers, A.~W.}, \bibinfo{author}{Campbell, C.~T.} \&
  \bibinfo{author}{Whiteoak, J.~B.}
\newblock \bibinfo{journal}{\bibinfo{title}{{A Catalogue of H {$\alpha$}
  -Emission Regions in the Southern Milky Way}}}.
\newblock {\emph{\JournalTitle{Monthly Notices of the Royal Astronomical
  Society}}} \textbf{\bibinfo{volume}{121}}, \bibinfo{pages}{103--110},
  \doiprefix\url{10.1093/MNRAS/121.1.103} (\bibinfo{year}{1960}).

\bibitem{Abdollahi2022IncrementalCatalog}
\bibinfo{author}{Abdollahi, S.} \emph{et~al.}
\newblock \bibinfo{journal}{\bibinfo{title}{{Incremental Fermi Large Area
  Telescope Fourth Source Catalog}}}.
\newblock {\emph{\JournalTitle{The Astrophysical Journal Supplement Series}}}
  \textbf{\bibinfo{volume}{260}}, \doiprefix\url{10.3847/1538-4365/ac6751}
  (\bibinfo{year}{2022}).

\bibitem{Manchester2005TheCatalogue}
\bibinfo{author}{Manchester, R.~N.}, \bibinfo{author}{Hobbs, G.~B.},
  \bibinfo{author}{Teoh, A.} \& \bibinfo{author}{Hobbs, M.}
\newblock \bibinfo{journal}{\bibinfo{title}{{The Australia Telescope National
  Facility Pulsar Catalogue}}}.
\newblock {\emph{\JournalTitle{The Astronomical Journal}}}
  \textbf{\bibinfo{volume}{129}}, \doiprefix\url{10.1086/428488}
  (\bibinfo{year}{2005}).

\bibitem{Malyshev2023Multi-classDefinition}
\bibinfo{author}{Malyshev, D.~V.} \& \bibinfo{author}{Bhat, A.}
\newblock \bibinfo{journal}{\bibinfo{title}{{Multi-class classification of
  Fermi-LAT sources with hierarchical class definition}}}.
\newblock {\emph{\JournalTitle{MNRAS}}} \textbf{\bibinfo{volume}{521}},
  \bibinfo{pages}{6195--6209}, \doiprefix\url{10.1093/mnras/stad940}
  (\bibinfo{year}{2023}).

\bibitem{Wolk2008HandbookRegions}
\bibinfo{author}{Wolk, S.~J.}, \bibinfo{author}{Bourke, T.~L.} \&
  \bibinfo{author}{Vigil, M.}
\newblock \emph{\bibinfo{title}{{Handbook of Star Forming Regions}}},
  vol.~\bibinfo{volume}{II} (\bibinfo{publisher}{Astronomical Society of the
  Pacific}, \bibinfo{year}{2008}).

\bibitem{Wolk2002DISCOVERY38}
\bibinfo{author}{Wolk, S.~J.}, \bibinfo{author}{Bourke, T.~L.},
  \bibinfo{author}{Smith, R.~K.}, \bibinfo{author}{Spitzbart, B.} \&
  \bibinfo{author}{Alves, J.~O.}
\newblock \bibinfo{journal}{\bibinfo{title}{{Discovery of Nonthermal X-Ray
  Emission from the Embedded Massive Star-forming Region RCW 38}}}.
\newblock {\emph{\JournalTitle{The Astrophysical Journal}}}
  \textbf{\bibinfo{volume}{580}}, \bibinfo{pages}{161--165},
  \doiprefix\url{10.1086/345611} (\bibinfo{year}{2002}).

\bibitem{Zucker2020AHandbook}
\bibinfo{author}{Zucker, C.} \emph{et~al.}
\newblock \bibinfo{journal}{\bibinfo{title}{{A compendium of distances to
  molecular clouds in the Star Formation Handbook}}}.
\newblock {\emph{\JournalTitle{Astronomy {\&} Astrophysics}}}
  \textbf{\bibinfo{volume}{633}}, \bibinfo{pages}{A51},
  \doiprefix\url{10.1051/0004-6361/201936145} (\bibinfo{year}{2020}).

\bibitem{Ellerbroek2013Rcw36:Formation}
\bibinfo{author}{Ellerbroek, L.~E.} \emph{et~al.}
\newblock \bibinfo{journal}{\bibinfo{title}{{Rcw36: Characterizing the outcome
  of massive star formation}}}.
\newblock {\emph{\JournalTitle{Astronomy and Astrophysics}}}
  \textbf{\bibinfo{volume}{558}}, \doiprefix\url{10.1051/0004-6361/201321752}
  (\bibinfo{year}{2013}).

\bibitem{Prisinzano2018AstrophysicsStars}
\bibinfo{author}{Prisinzano, L.} \emph{et~al.}
\newblock \bibinfo{journal}{\bibinfo{title}{{Astrophysics Low-mass star
  formation and subclustering in the H II regions RCW 32, 33, and 27 of the
  Vela Molecular Ridge A photometric diagnostics for identifying M-type
  stars}}}.
\newblock {\emph{\JournalTitle{Astronomy {\&} Astrophysics}}}
  \textbf{\bibinfo{volume}{617}}, \bibinfo{pages}{63},
  \doiprefix\url{10.1051/0004-6361/201833172} (\bibinfo{year}{2018}).

\bibitem{Massi2003StarC-cloud}
\bibinfo{author}{Massi, F.}, \bibinfo{author}{Lorenzetti, D.} \&
  \bibinfo{author}{Giannini, T.}
\newblock \bibinfo{journal}{\bibinfo{title}{{Star formation in the Vela
  molecular clouds. V. Young stellar objects and star clusters towards the
  C-cloud}}}.
\newblock {\emph{\JournalTitle{Astronomy and Astrophysics, v.399, p.147-167
  (2003)}}} \textbf{\bibinfo{volume}{399}}, \bibinfo{pages}{147},
  \doiprefix\url{10.1051/0004-6361:20021717} (\bibinfo{year}{2003}).

\bibitem{Bik2010SequentialRegions}
\bibinfo{author}{Bik, A.} \emph{et~al.}
\newblock \bibinfo{journal}{\bibinfo{title}{{Sequential star formation in RCW
  34: A spectroscopic census of the stellar content of high-mass star-forming
  regions}}}.
\newblock {\emph{\JournalTitle{Astrophysical Journal}}}
  \textbf{\bibinfo{volume}{713}}, \doiprefix\url{10.1088/0004-637X/713/2/883}
  (\bibinfo{year}{2010}).

\bibitem{Weaver1977InterstellarEvolution.}
\bibinfo{author}{Weaver, R.} \emph{et~al.}
\newblock \bibinfo{journal}{\bibinfo{title}{{Interstellar bubbles. II.
  Structure and evolution.}}}
\newblock {\emph{\JournalTitle{ApJ}}} \textbf{\bibinfo{volume}{218}},
  \bibinfo{pages}{377--395}, \doiprefix\url{10.1086/155692}
  (\bibinfo{year}{1977}).

\bibitem{Lallement2022UpdatedDust}
\bibinfo{author}{Lallement, R.}, \bibinfo{author}{Vergely, J.~L.},
  \bibinfo{author}{Babusiaux, C.} \& \bibinfo{author}{Cox, N.~L.}
\newblock \bibinfo{journal}{\bibinfo{title}{{Updated Gaia -2MASS 3D maps of
  Galactic interstellar dust}}}.
\newblock {\emph{\JournalTitle{Astronomy and Astrophysics}}}
  \textbf{\bibinfo{volume}{661}}, \doiprefix\url{10.1051/0004-6361/202142846}
  (\bibinfo{year}{2022}).

\bibitem{Yadav2017HowSuperbubbles}
\bibinfo{author}{Yadav, N.}, \bibinfo{author}{Mukherjee, D.},
  \bibinfo{author}{Sharma, P.} \& \bibinfo{author}{Nath, B.~B.}
\newblock \bibinfo{journal}{\bibinfo{title}{{How multiple supernovae overlap to
  form superbubbles}}}.
\newblock {\emph{\JournalTitle{Monthly Notices of the Royal Astronomical
  Society}}} \textbf{\bibinfo{volume}{465}}, \bibinfo{pages}{1720--1740},
  \doiprefix\url{10.1093/MNRAS/STW2522} (\bibinfo{year}{2017}).

\bibitem{Canto2000THESTARS}
\bibinfo{author}{Cant{\`{o}}, J.}, \bibinfo{author}{Raga, A.~C.} \&
  \bibinfo{author}{Rodriguez, L.~F.}
\newblock \bibinfo{journal}{\bibinfo{title}{{The Hot, Diffuse Gas in a Dense
  Cluster of Massive Stars}}}.
\newblock {\emph{\JournalTitle{The Astrophysical Journal}}}
  \textbf{\bibinfo{volume}{536}}, \bibinfo{pages}{896--901},
  \doiprefix\url{10.1086/308983} (\bibinfo{year}{2000}).

\bibitem{Bourke2001NEWCLOUDS}
\bibinfo{author}{Bourke, T.~L.}, \bibinfo{author}{Myers, P.~C.},
  \bibinfo{author}{Robinson, G.} \& \bibinfo{author}{Hyland, A.~R.}
\newblock \bibinfo{journal}{\bibinfo{title}{{New OH Zeeman Measurements of
  Magnetic Field Strengths in Molecular Clouds}}}.
\newblock {\emph{\JournalTitle{The Astrophysical Journal}}}
  \textbf{\bibinfo{volume}{554}}, \bibinfo{pages}{916--932},
  \doiprefix\url{10.1086/321405} (\bibinfo{year}{2001}).

\bibitem{Badmaev2022InsideSimulations}
\bibinfo{author}{{Badmaev}, D.~V.}, \bibinfo{author}{{Bykov}, A.~M.} \&
  \bibinfo{author}{{Kalyashova}, M.~E.}
\newblock \bibinfo{journal}{\bibinfo{title}{{Inside the core of a young massive
  star cluster: 3D MHD simulations}}}.
\newblock {\emph{\JournalTitle{mnras}}} \textbf{\bibinfo{volume}{517}},
  \bibinfo{pages}{2818--2830}, \doiprefix\url{10.1093/mnras/stac2738}
  (\bibinfo{year}{2022}).
\newblock \eprint{2209.11465}.

\bibitem{Padovani2019Non-thermalRegions}
\bibinfo{author}{{Padovani}, M.}, \bibinfo{author}{{Marcowith}, A.},
  \bibinfo{author}{{S{\'a}nchez-Monge}, {\'A}.}, \bibinfo{author}{{Meng}, F.}
  \& \bibinfo{author}{{Schilke}, P.}
\newblock \bibinfo{journal}{\bibinfo{title}{{Non-thermal emission from cosmic
  rays accelerated in H II regions}}}.
\newblock {\emph{\JournalTitle{aap}}} \textbf{\bibinfo{volume}{630}},
  \bibinfo{pages}{A72}, \doiprefix\url{10.1051/0004-6361/201935919}
  (\bibinfo{year}{2019}).
\newblock \eprint{1908.07246}.

\bibitem{Zabalza2015}
\bibinfo{author}{Zabalza, V.}
\newblock \bibinfo{title}{{Naima: A Python package for inference of
  relativistic particle energy distributions from observed nonthermal
  spectra}}.
\newblock In \emph{\bibinfo{booktitle}{Proceedings of Science}}, vol.
  \bibinfo{volume}{30-July-20} (\bibinfo{publisher}{Proceedings of Science
  (PoS)}, \bibinfo{year}{2015}).

\bibitem{Aharonian2004}
\bibinfo{author}{Aharonian, F.~A.}
\newblock \emph{\bibinfo{title}{{Very High Energy Cosmic Gamma Radiation - A
  Crucial Window on the Extreme Universe}}} (\bibinfo{publisher}{World
  Scientific Publishing Co. Pte. Ltd.}, \bibinfo{year}{2004}).

\bibitem{Tatischeff2021TheComposition}
\bibinfo{author}{Tatischeff, V.}, \bibinfo{author}{Raymond, J.~C.},
  \bibinfo{author}{Duprat, J.}, \bibinfo{author}{Gabici, S.} \&
  \bibinfo{author}{Recchia, S.}
\newblock \bibinfo{journal}{\bibinfo{title}{{The origin of Galactic cosmic rays
  as revealed by their composition}}}.
\newblock {\emph{\JournalTitle{Monthly Notices of the Royal Astronomical
  Society}}} \textbf{\bibinfo{volume}{508}}, \bibinfo{pages}{1321--1345},
  \doiprefix\url{10.1093/MNRAS/STAB2533} (\bibinfo{year}{2021}).

\bibitem{Massi2019DenseC}
\bibinfo{author}{Massi, F.} \emph{et~al.}
\newblock \bibinfo{journal}{\bibinfo{title}{{Dense cores and star formation in
  the giant molecular cloud Vela C}}}.
\newblock {\emph{\JournalTitle{Astronomy and Astrophysics}}}
  \textbf{\bibinfo{volume}{628}}, \bibinfo{pages}{A110},
  \doiprefix\url{10.1051/0004-6361/201935047} (\bibinfo{year}{2019}).

\bibitem{Dame2000}
\bibinfo{author}{Dame, T.~M.}, \bibinfo{author}{Hartmann, D.} \&
  \bibinfo{author}{Thaddeus, P.}
\newblock \bibinfo{journal}{\bibinfo{title}{{The Milky Way in Molecular Clouds:
  A New Complete CO Survey}}}.
\newblock {\emph{\JournalTitle{The Astrophysical Journal}}}
  \textbf{\bibinfo{volume}{547}}, \bibinfo{pages}{792--813},
  \doiprefix\url{10.1086/318388} (\bibinfo{year}{2000}).

\bibitem{BenBekhti2016}
\bibinfo{author}{Ben~Bekhti, N.} \emph{et~al.}
\newblock \bibinfo{journal}{\bibinfo{title}{{HI4PI: A full-sky Hi survey based
  on EBHIS and GASS}}}.
\newblock {\emph{\JournalTitle{Astronomy and Astrophysics}}}
  \textbf{\bibinfo{volume}{594}}, \doiprefix\url{10.1051/0004-6361/201629178}
  (\bibinfo{year}{2016}).

\bibitem{Ade2011}
\bibinfo{author}{Ade, P.}, \bibinfo{author}{Aghanim, N.},
  \bibinfo{author}{Arnaud, M.} \& \bibinfo{author}{Ashdown, M.}
\newblock \bibinfo{journal}{\bibinfo{title}{{Planck early results. XIX. All-sky
  temperature and dust optical depth from Planck and IRAS. Constraints on the
  “dark gas” in our Galaxy}}}.
\newblock {\emph{\JournalTitle{Astronomy \& Astrophysics}}}
  \textbf{\bibinfo{volume}{536}}, \bibinfo{pages}{16},
  \doiprefix\url{10.1051/0004-6361/201116479} (\bibinfo{year}{2011}).

\bibitem{Ackermann2017SearchGeV}
\bibinfo{author}{Ackermann, M.} \emph{et~al.}
\newblock \bibinfo{journal}{\bibinfo{title}{{Search for Extended Sources in the
  Galactic Plane Using Six Years of Fermi-Large Area Telescope Pass 8 Data
  above 10 GeV}}}.
\newblock {\emph{\JournalTitle{The Astrophysical Journal}}}
  \textbf{\bibinfo{volume}{843}}, \bibinfo{pages}{139},
  \doiprefix\url{10.3847/1538-4357/aa775a} (\bibinfo{year}{2017}).

\bibitem{Ackermann2017SearchGeVb}
\bibinfo{author}{Ackermann, M.} \emph{et~al.}
\newblock \bibinfo{journal}{\bibinfo{title}{{Search for Extended Sources in the
  Galactic Plane Using Six Years of Fermi-Large Area Telescope Pass 8 Data
  above 10 GeV}}}.
\newblock {\emph{\JournalTitle{The Astrophysical Journal}}}
  \textbf{\bibinfo{volume}{843}}, \bibinfo{pages}{139},
  \doiprefix\url{10.3847/1538-4357/AA775A} (\bibinfo{year}{2017}).

\bibitem{Mattox1996}
\bibinfo{author}{Mattox, J.~R.} \emph{et~al.}
\newblock \bibinfo{journal}{\bibinfo{title}{{The Likelihood Analysis of EGRET
  Data}}}.
\newblock {\emph{\JournalTitle{The Astrophysical Journal}}}
  \textbf{\bibinfo{volume}{461}}, \bibinfo{pages}{396},
  \doiprefix\url{10.1086/177068} (\bibinfo{year}{1996}).

\bibitem{TheFermi-LATcollaboration2019}
\bibinfo{author}{{The Fermi-LAT collaboration}}.
\newblock \bibinfo{journal}{\bibinfo{title}{{Fermi Large Area Telescope Fourth
  Source Catalog}}}.
\newblock {\emph{\JournalTitle{The Astrophysical Journal Supplement Series}}}
  \textbf{\bibinfo{volume}{242}}, \doiprefix\url{10.3847/1538-4365/ab6bcb}
  (\bibinfo{year}{2019}).

\bibitem{Vladimirov2011}
\bibinfo{author}{Vladimirov, A.~E.} \emph{et~al.}
\newblock \bibinfo{journal}{\bibinfo{title}{{GALPROP WebRun: An internet-based
  service for calculating galactic cosmic ray propagation and associated photon
  emissions}}}.
\newblock {\emph{\JournalTitle{Computer Physics Communications}}}
  \textbf{\bibinfo{volume}{182}}, \bibinfo{pages}{1156--1161},
  \doiprefix\url{10.1016/j.cpc.2011.01.017} (\bibinfo{year}{2011}).

\bibitem{Yusifov2004}
\bibinfo{author}{Yusifov, I.} \& \bibinfo{author}{Kucuk, I.}
\newblock \bibinfo{journal}{\bibinfo{title}{{Revisiting the radial distribution
  of pulsars in the Galaxy}}}.
\newblock {\emph{\JournalTitle{Astronomy and Astrophysics}}}
  \textbf{\bibinfo{volume}{422}}, \bibinfo{pages}{545--553},
  \doiprefix\url{10.1051/0004-6361:20040152} (\bibinfo{year}{2004}).

\bibitem{Ballet2023}
\bibinfo{author}{{Ballet}, J.}, \bibinfo{author}{{Bruel}, P.},
  \bibinfo{author}{{Burnett}, T.~H.}, \bibinfo{author}{{Lott}, B.} \&
  \bibinfo{author}{{The Fermi-LAT collaboration}}.
\newblock \bibinfo{journal}{\bibinfo{title}{{Fermi Large Area Telescope Fourth
  Source Catalog Data Release 4 (4FGL-DR4)}}}.
\newblock {\emph{\JournalTitle{arXiv e-prints}}}
  \bibinfo{pages}{arXiv:2307.12546}, \doiprefix\url{10.48550/arXiv.2307.12546}
  (\bibinfo{year}{2023}).
\newblock \eprint{2307.12546}.

\bibitem{Bolatto2013}
\bibinfo{author}{Bolatto, A.~D.}, \bibinfo{author}{Wolfire, M.} \&
  \bibinfo{author}{Leroy, A.~K.}
\newblock \bibinfo{journal}{\bibinfo{title}{{ The CO-to-H 2 Conversion Factor
  }}}.
\newblock {\emph{\JournalTitle{Annual Review of Astronomy and Astrophysics}}}
  \textbf{\bibinfo{volume}{51}}, \bibinfo{pages}{207--268},
  \doiprefix\url{10.1146/annurev-astro-082812-140944} (\bibinfo{year}{2013}).

\bibitem{Acero2016}
\bibinfo{author}{Acero, F.} \emph{et~al.}
\newblock \bibinfo{journal}{\bibinfo{title}{{Development of the Model of
  Galactic Interstellar Emission for Standard Point-Source Analysis of Fermi
  Large Area Telescope Data}}}.
\newblock {\emph{\JournalTitle{The Astrophysical Journal Supplement Series}}}
  \textbf{\bibinfo{volume}{223}}, \doiprefix\url{10.3847/0067-0049/223/2/26}
  (\bibinfo{year}{2016}).

\bibitem{Park2015SimultaneousShocks}
\bibinfo{author}{Park, J.}, \bibinfo{author}{Caprioli, D.} \&
  \bibinfo{author}{Spitkovsky, A.}
\newblock \bibinfo{journal}{\bibinfo{title}{{Simultaneous acceleration of
  protons and electrons at nonrelativistic quasiparallel collisionless
  shocks}}}.
\newblock {\emph{\JournalTitle{Physical Review Letters}}}
  \textbf{\bibinfo{volume}{114}}, \bibinfo{pages}{085003},
  \doiprefix\url{https://doi.org/10.1103/PhysRevLett.114.085003}
  (\bibinfo{year}{2015}).

\bibitem{Kafexhiu2014}
\bibinfo{author}{Kafexhiu, E.}, \bibinfo{author}{Aharonian, F.},
  \bibinfo{author}{Taylor, A.~M.} \& \bibinfo{author}{Vila, G.~S.}
\newblock \bibinfo{journal}{\bibinfo{title}{{Parametrization of gamma-ray
  production cross-sections for pp interactions in a broad proton energy range
  from the kinematic threshold to PeV energies}}}.
\newblock {\emph{\JournalTitle{Physical Review D}}}
  \textbf{\bibinfo{volume}{90}}, \bibinfo{pages}{123014}
  (\bibinfo{year}{2014}).

\end{thebibliography}

\end{document}